# A Universe without Dark Energy and Dark Matter

Shlomo Barak and Elia M. Leibowitz

School of Physics & Astronomy, Tel Aviv University Tel Aviv, 69978, Israel

arxiv:astro-ph                                    26 March 2010

## Abstract

The universe has evolved to be a filamentary web of galaxies and large inter-galactic zones of space without matter. The Euclidian nature of the universe indicates that it is not a 3D manifold within space with an extra spatial dimension. This justifies our assumption that the FRW space-time evolves in the inter-galactic zones like separate FRW universes. Thus we do not necessarily have to consider the entirety of the universe. Our assumption enables us to prove that:

- In the current epoch, space in the intergalactic zones expands at a constant rate.
- In and around galaxies, space expansion is inhibited.

With these results, and an extended Gauss Theorem for a deformed space, we show that there is no need for the hypothetical Dark Energy (DE) and Dark Matter (DM) to explain phenomena attributed to them.

Key words: cosmic microwave background, dark energy, dark matter, gravitation, relativity.

Acronyms used are listed at the end of this paper.

## 1.     Introduction

### 1.1    Our Idea

Observations show that our universe is Euclidian (flat) i.e., its curvature is $k = 0$. In current Cosmology, a universe with positive curvature is modeled as a three-sphere. For $k = 0$, the radius of the universe in the fourth spatial dimension is $r \rightarrow \infty$ (B. F. Schutz, 2003). This is equivalent to a 3D universe as part of a 3D elastic space with no extra spatial dimension. This leads to our assumption that we can explore the Inter-galactic Zones (IZ) with no matter, separately from the much smaller Zones with Matter (MZ) and avoid dealing with the entirety of the universe (E. Bertschinger, 2006).

We show that the combined volume of IZs is $10^3$ larger than that of the combined MZs. By using the second Friedmann equation we show that space in IZs expands, in this epoch, at a constant rate, whereas space in MZs does not expand. We also examine the border between an IZ and a MZ and show that the rate of expansion here is gradually reduced to zero. This border is created where, and when, a balance of positive and negative pressures exists.

The positive pressure in our discussion is the local Gravitational Field Energy (GFE) density. The negative pressure, as we show, is the Cosmological Microwave Background (CMB) energy density.

In General Relativity (GR), the nature of the GFE density $\in_g$ is problematic. However, we consider it as the GFE density in the cosmological frame of reference, in which the CMB is isotropic.

GR considers CMB energy density, $\in_{CMB}$, being ElectroMagnetic (EM) energy density, as having a positive pressure. However, we show that $\in_{CMB}$ is a negative pressure and that this does not contradict GR.

Note that the current understanding is that space in our universe expands homogeneously and isotropically, everywhere (Davis et al 2003, 2007). It is further understood that material bodies are held together mainly by gravitation and EM forces, and are not affected by space expansion. However, according to GR, gravitation is the contraction (curving) of space around masses. Hence it is reasonable to consider the possibility that this contraction affects space around galaxies by causing, locally, the expansion to be inhomogeneous and anisotropic.

Since MZs are zones of non-expanding space, they do not contribute to red-shifting. Light from a galaxy with a given redshift, z, passes through both IZs and MZs. Hence the galaxy is located at a larger distance (about 10 %) than z indicates. This is why supernovae appear fainter than expected. In this approach <u>there is no need for DE</u>.

Phenomena attributed to DM are proven to be the result of strong deformations of space, in and around galaxies – the MZs. These deformations are created by space contraction by matter and opposing expansion (space dilation).

In a homogeneous and isotropic space, the gravitational field is determined by the flux generated by a mass M, as Gauss Theorem shows. However, in a deformed space this theorem must be extended, to account for the foamy structure of space. As a result, the expression for the gravitational field is no longer Newtonian.

We derive, using our extended Gauss Theorem, the dynamic and kinematic (Tully-Fisher) relations that govern the motions of celestial bodies in and around galaxies and show how they account for the flattening of Rotation Curves (RC). In this approach <u>there is no need for DM</u>.

### 1.2  Outline

A. We assume that, since the universe is Euclidian, we can relate to IZs as separate FRW universes.

B. We assume that the $\in_{CMB}$ is a negative pressure. This is in contrast to current understanding that it is a positive pressure like $\in_g$.

C. Based on Steps A and B, we show that for IZs in this epoch $\ddot{a}$, the second derivative of the Cosmological Scale Factor (CSF) a, is positive but approximately zero. Hence $\dot{a} = \text{const}$.

D. Based on Steps A and B, we show that for MZs both $\ddot{a}$ and $\dot{a}$ are zero. Thus, in MZs, space does not expand and, of necessity, cannot contribute to red shifting.

E. We estimate the ratio of the combined volume of IZs to that of MZs to be $10^3$, hence the path through MZs is ~0.1 of the path through IZs. This impacts the derivation and calculation of the luminosity distance.

F. We conclude, based on Steps C and D, that the border between an IZ and a MZ is set where, and when, $|\epsilon_g| = \epsilon_{CMB}$. Inside MZs $|\epsilon_g| > \epsilon_{CMB}$. At a later stage, we show that the MZs are the DM halos.

G. We derive $H(z) = H_0 h(z)$, based on Step C.

H. We derive $d_L$, the Luminosity Distance (LD), based on Steps C and D.

I. We plot the curve of μ, the Distance Modulus (DM) as a function of log z, based on Step H. We calculate $\chi^2$ and show that our fit to data from supernovae is as good as that of ΛCDM.

J. Based on I, we calculate the ratio of the combined volume of the IZs to that of the MZs and get ~$10^3$ (our estimation in Step E).

K. We develop an extended Gauss Theorem for the case of a deformed space. This extension is based on the idea that space is foamy.

L. We derive, based on Step K, the <u>dynamic</u> Extended Newtonian Gravitational Law and <u>kinematic</u> Tully-Fisher relations that govern the motions of celestial bodies in and around galaxies. This derivation does not require any gravitating matter beyond the observed baryonic matter.

M. We show that the theoretical RCs resulting from the relations in Step L fit observed RCs.

N. We dispel the need for DE and DM, based on Steps I, L and M.

O. Our results justify the assumptions in Steps A and B.

## 1.3  DE and DM – a Historical Note

DE was suggested to explain the <u>supposed</u> changes in the rate of expansion of the universe, and is also used, together with DM, to explain its Euclidian nature (flatness), (Riess et al 1998, Perlmutter et al 1999). However, DE has never been conclusively identified or directly detected.

DM was suggested, based on Newtonian Physics, by Oort (1932) and Zwicky (1933) to explain the seemingly non-Newtonian dynamics within the Milky Way galaxy and in clusters of galaxies. In the seventies, the discovery of flat RCs in and around galaxies (Rubin and Ford 1970; Ostriker, Peebles and Yahil 1974) added support for the idea of DM. However, DM has never been conclusively identified nor directly detected.

## 1.4  Remarks

Λ, the Cosmological Constant, does not appear in this paper.

In this paper, the pressure, p, is related to energy density $\epsilon$ as $3p = \epsilon$. Both p and $\epsilon$ can be positive or negative. It is also possible that $3p = \pm|\epsilon|$.

For more on the web-like structure of the universe see R. Van De Weygaert and E. Platen (2009).

## 2. The Expansion of the Universe

### 2.1 The Expansion of IZs

The known Friedmann equation for a "perfect fluid" universe, equation 18.10, Page 393, in the textbook *Relativity* by Rindler (2004) is.

(1) $\quad \ddot{a} = -\dfrac{4\pi G}{3}\left(m + \dfrac{3p}{c^2}\right)a$

where a is the CSF, m is the mass density of the universe and p is pressure.

For IZs, being zones without matter, we take m = 0. The $p_g = 1/3\left|\in_g\right|$ and $p_{CMB} = -1/3 \in_{CMB}$ are the main (if not the only) contributors to p.

(2) $\quad p = p_g + p_{CMB}$

The $\in_g$ in IZs is zero, except on their boundaries (close to masses). Note that from symmetry considerations, or the fact that k = 0, the gravitational field in IZs, far from masses is zero, hence $\in_g$ is zero. However, close to masses, $\in_g \neq 0$. The $\in_g$ is thus a local attribute and is considered as being a positive pressure (contraction).

GR considers the energy of EM waves as having the same positive contribution to curving as that of matter, and the same type of pressure for both $\in_{CMB}$ and $\in_g$.

In contrast to this current thinking, we contend that the energy density of EM waves is a negative pressure and as such contributes to the negative curving of space (dilating it). Note that this is the kind of curving currently attributed to DE.

Our contention seems to contradict GR since it can be wrongly understood as implying that a beam of light bends away from a mass rather than towards it. This understanding arises from the belief that photons are <u>independent</u> particles and, as such, a negative curvature contribution of the energy of photons would imply an anti-gravitational equivalent mass. However, our suggestion does <u>not</u> mean that light bends away from a mass – clearly light bends towards a mass, as experience shows.

The situation is clarified if photons are considered, as they should be, as wavepackets and not as independent particles (R. Loudon, 2000). Their velocity is determined by the permittivity and permeability of space in their tracks, which are affected by the presence of a large mass. Hence, they bend towards the mass despite their individual negative contributions to the curvature of space, which is negligible.

We prove our contention, Section 8, by obtaining the correct dynamic and kinematic relations that govern the motions of celestial bodies in and around galaxies. Thus we attribute a negative sign to $\in_{CMB}$ in the substitution for $3p = \in$ in equation (1). We get:

(3) $\quad \ddot{a} = \dfrac{4\pi G}{3}\dfrac{\in_{CMB}}{c^2}a$

and thus it is the $\epsilon_{CMB}$ that is responsible for an accelerated expansion. However, we are interested in $\ddot{a}$ in the current epoch for which we have data for the light coming from supernovae. This epoch is the range for z between 0 and 4.

In the ΛCDM model, cosmologists relate to ~0.71 of the critical value $m_{crit}$ ~ $10^{-29}$ gr cm$^{-3}$ as the contribution of DE, in the calculation of $\ddot{a}$. However, in equation (3) $\frac{\epsilon_{CMB}(today)}{c^2} \sim 4 \cdot 10^{-34}$ gm cm$^{-3}$. This difference in the orders of magnitude justifies the approximation for our epoch:

(4) $\quad \ddot{a} = 0 \quad$ and hence:

(5) $\quad \dot{a} = \text{const} \quad$ which is the rate of expansion

Note that the volume of the non-expanding zones of space – MZs (Dark Matter Halos) around galaxies, see Section 3.2, is only a small fraction of the total volume of an Hubble Sphere (HS). This fraction is approximately $10^{-3}$. This estimation is obtained by taking the average radius of a MZ as 200 KPC, and the number of galaxies as $250 \cdot 10^9$.

The fraction of the combined linear dimensions of MZs to that of IZs is thus $(10^{-3})^{\frac{1}{3}} = 0.1$. We also show that an MZ is enlarged in step with a. However, in Section 5.3 we obtain a similar, calculated, result for the above fraction.

## 2.2　No Expansion of MZs

In and around MZs, on the boundary of the IZs, $\epsilon_g \neq 0$.

The Friedmann equation (1) for the boundary becomes:

$$\frac{\ddot{a}}{a} = -\frac{4\pi G}{3c^2}(3p_g + 3p_{CMB}) = -\frac{4\pi G}{3c^2}(|\epsilon_g| - \epsilon_{CMB}) \quad \text{or:}$$

(6) $\quad \frac{\ddot{a}}{a} = \frac{4\pi G}{3c^2}(\epsilon_{CMB} - |\epsilon_g|)$

Thus space expands only where and when $\epsilon_{CMB} \geq |\epsilon_g|$. At a distance R, from the center of a galaxy, where $\epsilon_{CMB} = |\epsilon_g|$, we get $\ddot{a} = 0$, but for r < R we get $\ddot{a} < 0$ (since $\epsilon_{CMB} < |\epsilon_g|$) and $\dot{a}$ drops to zero.

Note that, if, at R, space expands, space anywhere inside a sphere with radius less than R, cannot contract. Thus the rate of expansion can only be reduced to $\dot{a} = 0$ (cannot be negative). Let the distance, $R_0$, from the center of the galaxy, be the distance at which, initially, at the time of galaxy formation, $\dot{a}$ was zero. With time, $\dot{a} = 0$ is reached at a larger distance, R, since $\epsilon_{CMB}$ falls with the expansion. The radius of this sphere of non-expanding space grows with time from $R_0$ to R - this is a DM halo, as explained in Section 8.

In this evolving non-expanding zone around galaxies, which we designate as MZ, space close to the galaxy is denser (lower a) than space at a larger distance (higher a). The expansion of the universe is thus inhomogeneous and anisotropic.

## 3. The Fit of Our Model to Observations

### 3.1 Our Luminosity Distance (LD) Ignoring the MZs

$H \stackrel{def}{=} \dot{a}/a$, but, as we have shown: $\dot{a} = $ const ($\ddot{a} = 0$). Hence, using the notation $\dot{a} = H_0$ gives:
$H = H_0 a^{-1}$.

The value of the constant $\dot{a}$ is $2.3 \cdot 10^{-18}$ sec$^{-1}$, since $H_0 = \dot{a}/a_0$, where $a_0 = 1$ today.

The relations $H = H_0 a^{-1}$ and $a = 1/(1+z)$ give $H(z) = H_0(1 + z)$, from which we derive the LD notated $d_L$.

In this section we use the conventional notation $H(z) = H_0 h(z)$. In <u>our</u> theory:

(7)  $h(z) = 1 + z$

Whereas the <u>known</u> equation with the two dependent free parameters, $\Omega_M$ and $\Omega_\Lambda$, for <u>flat space</u> where $\Omega_M + \Omega_\Lambda = 1$ is:

(8)  $h(z) = \left[(1+z)^2(1+\Omega_M z) - z(2+z)\Omega_\Lambda\right]^{\frac{1}{2}}$   (Perlmutter, 1997).

Note that our h(z), equation (7), is identical to the h(z) in (8) if $\Omega_M = \Omega_\Lambda = 0$.

In the literature (Perlmutter and Schmidt, 2003, E. W. Kolb, 2007), the level of confidence in LD based on h(z), Equation (8) with $\Omega_M = \Omega_\Lambda = 0$ is low compared to that for $\Omega_M = 0.3$ and $\Omega_\Lambda = 0.7$. A more realistic LD, $d_L$, is obtained by taking into account the path through MZs, ("DM halos"), which in Section 2.2 are proven to be zones of non-expanding space and hence do not give rise to redshift. This $d_L$ gives a fit which is as good as the $\Lambda$CDM fit.

This issue is discussed in the Section 3.2.

Our h(z) yields a different $d_L$ from that derived from equation (8). LD is defined by the ratio of the luminosity, L, of a supernova, to its measured flux, F:

(9)  $d_L^2 \equiv \dfrac{L}{4\pi F}$

From the known relation:

(10)  $d_L = (1+z)\dfrac{c}{H_0} \cdot \int_0^z \dfrac{dz'}{h(z')}$   using <u>our</u> h(z), we get:

(11)  $d_L = (1+z)\dfrac{c}{H_0} \cdot \int_0^z \dfrac{dz'}{1+z'} = (1+z)\dfrac{c}{H_0} \ln(1+z)$

This LD (11) has already appeared in the literature although it has been derived from different cosmological models. However, the fit to data in this case is poorer than that of the $\Lambda$CDM. The criterion for fit is $\chi^2$, and in this case it is 10 times larger that that for the $\Lambda$CDM.

## 3.2 Our Corrected Luminosity Distance (LD) and Distance Modulus Taking MZs into Account

We present the corrected $d_L$ and show that its fit is as good as the fit obtained by the ΛCDM model, having the same $\chi^2$.

This section is based on results that appear in Section 8, which shows that MZs are the DM Halos, where space expansion is inhibited (frozen).

Light emitted from a galaxy, on its way to us, passes through both IZs and MZs. The expanding IZs contribute to cosmological red-shifting, whereas the non-expanding MZs do not. The LD to a galaxy with red-shift z is thus composed of contributions from IZs and MZs. These contributions are $D_I$ and $D_M$ respectively. Hence:

(12) $\quad d_L(z) = (1+z)(D_I + D_M)$

(13) $\quad D_I(z) = \dfrac{c}{H_0} \ln(1+z) \qquad$ See (11).

We now derive the term $D_M$ for MZs and incorporate it in (12) for $d_L(z)$.

We assume that the majority of galaxies were formed during the same epoch, designated by the time $t_0$. At $t_0$ the scale factor was $a_0$, and the MZs had spherical cores with radius $r_0$. Section 8 shows that the radius of a MZ grows with time. It shows that the density in the core of the galaxy is proportional to $a_0^{-3}$, and drops towards the outer edge of the MZ, where it is proportional to $a^{-3}$.

Let R(t) be the mean distance between us and an emitting galaxy at cosmic time t.

(14) $\quad R(t) = R_I(t) + R_M(t)$

where $R_I(t)$ is the sum of the distances though IZs and $R_M(t)$ is the sum of the distances through MZs.

In our universe, at the present cosmic time T: $\quad R(T) = R_I(T) + R_M(T)$.

$R_M(t)$, at each epoch z, depends on the number of MZ spheres and their radii.

To simplify our calculation, we model the universe as if all MZs were formed at z = 1,5 and with the same core radius. We thus represent all MZs by just one halo with the radius that is the sum of the radii of all spheres.

Our analysis of the DM phenomenon, Section 8, for the approximation of galaxies with a fixed mass M (no accretion or ejection of mass after formation) shows that the radius, $R_M(t)$, of the frozen halo at time $t > t_0$ is related to the core radius $r_0$ as:

(15) $\quad R_M(t) = r(t) = r_0 \left[ \dfrac{a(t)}{a_0} \right]^2 \qquad$ See (50).

We now define β as the ratio of the radius of the frozen sphere to the mean distance between us and the emitting galaxy, at the present epoch:

$$\text{(16)} \quad \beta = \frac{R_M(T)}{R(T)} = \frac{r(T)}{R(T)} = \frac{r_0\left(\frac{a(T)}{a_0}\right)^2}{R_I(T)+R_M(T)}$$

Using this definition we derive (Appendix A) the expression for the <u>corrected</u> distance modulus.

$$\text{(21)} \quad \mu(z) = 5 \cdot \left\{ \log\left[\left(1+1.5\beta\left[1-\frac{1}{(1+z)^2}\right]\right) \ln(1+z)\right] + \log(1+z) \right\} + \text{const}$$

With this expression we find for CONSTITUTION:

$\beta = 0.1104 \quad \mu_0 = 43.312 \quad \chi^2 = 466.94$.

For comparison, the parameters $(\Omega_m, 1-\Omega_m)$ which best fit the ΛCDM model, are:

$\Omega_m = 0.2902 \quad \mu_0 = 43.316 \quad \chi^2 = 465.52$

Figure 1 is a plot of the Distance Modulus μ of the 397 SN Ia objects of the Constitution data set of Hicken et al (2009), versus $\log_{10}$ of their corresponding redshifts, z.

The red curve is the best fit of our theoretical μ, given by (21).

The black curve is the best fit of the relations predicted by the ΛCDM theory for a flat universe, based on the LD equation presented in Perlmutter et al (1997).

Note that, for low z, the two curves coincide. However, for $\log_{10}(z) = 0.4$, i.e., z = 2.5, they diverge.

Frame (a) contains the full data set.

Frame (b) is a zoom on the sub-set of lower z.

Frame (c) is a zoom on the sub-set of higher z.

Frame (d) is an extension of the two theoretical curves up to the value z = 100.

Note that in Section 2.1 we have estimated the ratio of the combined radii of frozen spherical halos (MZs) to the mean intergalactic distance in the local universe. This estimation stemmed from our DM theory and the commonly accepted estimates of the number of galaxies in the universe. Our estimation results in a ratio of ~ 0.1 (the volumetric ratio of MZs to IZs is $10^{-3}$). The very same value, which is β, is obtained as a result of the best fitting of (21) to the SN Ia Constitution file.

The functional form of the coefficient $\left(1-\frac{1}{(1+z)^2}\right)$ in (21) is obtained from the general consideration resulting from our DM and DE theory. It turns out that this expression represents the curve that is best fitted to the observed data, as judged by the $\chi^2$ value. For example, the $\chi^2$ value that is obtained with an expression of the form $\left(1-\frac{1}{(1+z)^\alpha}\right)$ takes the smallest value for α = 2.

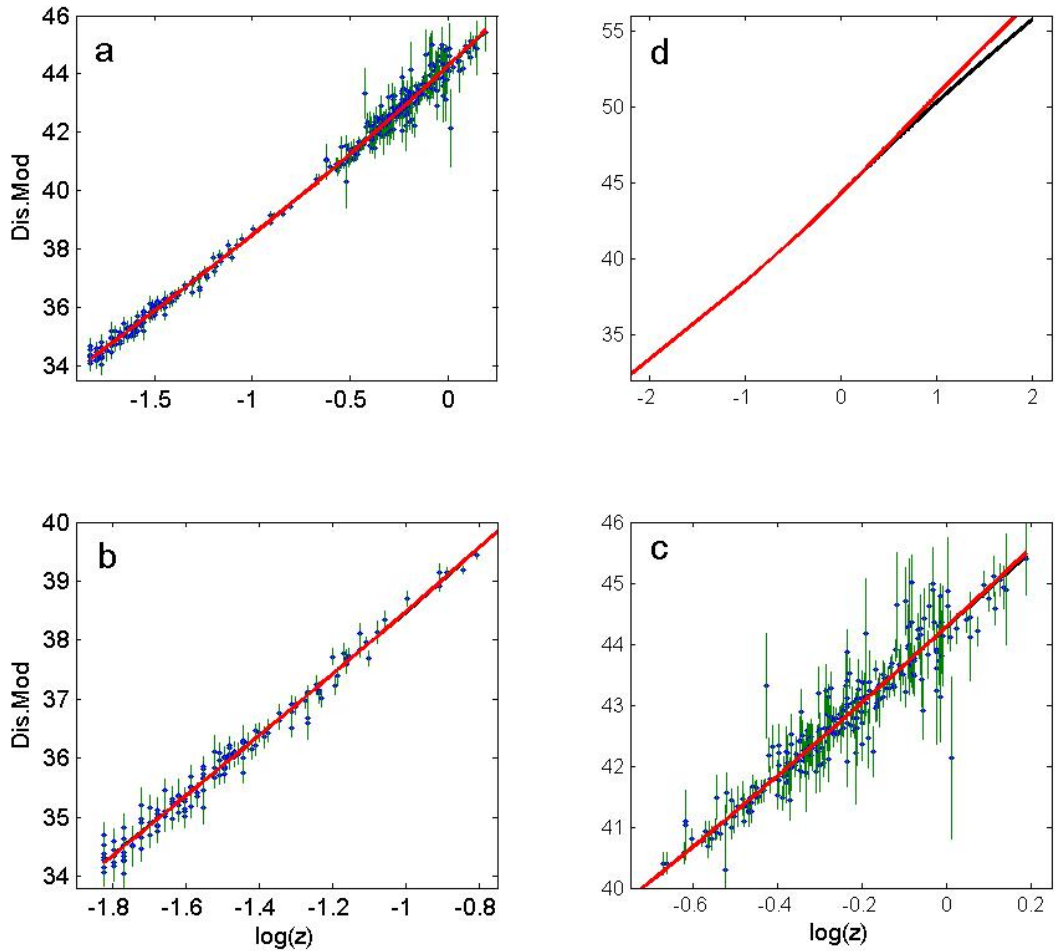

**Figure 1. Distance Modulus, μ, Versus the Log of the Redshift, z with Data Points for 397 Ia Supernovae from the Constitution**

## 4. Notes on Space and the Universe

### 4.1 Space is Foamy

The consensus that space is foamy, and hence cellular, rests on the meaning of expansion, and the requirement that its vibrations have a finite energy density. By "its vibration" we mean the ElectroMagnetic (EM) waves - this understanding is not crucial to our discussion and appears simply as a remark. The cut-off wavelength of the Zero Point Fluctuations (ZPF), which determines its energy density, is the smallest linear dimension of a space cell. Whether this linear dimension is Planck's length, or not, is not relevant to our discussion.

It is interesting to note that B. Riemann, quoted by Chandrasekhar in Nature (1990), was of the opinion that space is foamy.

### 4.2 Space is Deformable and Three-Dimensional

Our 3D space is <u>not</u> a curved 3D manifold in a hyperspace with an additional spatial dimension. However, for a globally flat universe the issue of an additional spatial dimension is not relevant. The terms "deformed" and "curved" are used for a 3D elastic space and a 3D-manifold, respectively. Note that Riemannian geometry is the geometry of both curved manifolds and deformed spaces. This is explained by A. Einstein (1921) and R. Feynman (1963).

The deformation of space is the change in size of its cells. Positive or negative deformation, around a point in space, means that the space cells grow or shrink, respectively, from this point outwards. For a positively curved manifold, the ratio of the circumference of a circle to the radius is less than $2\pi$, as measured by a rigid yardstick. For a deformed 3D-space, with a positive deformation, around a point the above ratio is also less than $2\pi$, as measured by a flexible yardstick such as the linear dimension of a space cell. Note that for a <u>deformed space</u> there is no meaning to global deformation, <u>deformation is a local attribute</u>. The surface of a sphere with radius R is a 2D manifold with a global Gaussian curvature $1/R^2$. However, a global homogeneous deformation for a deformable 2D planar sheet can only have the value zero i.e., the sheet is Euclidian. Here, deformation around a point is expressed by a scale factor $a(r,t)$ that depends on both time and the vector, r, from the point. Space density is proportional to $a(r,t)^{-3}$.

### 4.3 On Our Universe

For a 3D space manifold, curved in a hyperspace with an extra spatial dimension, the Cosmological Principle (CP) implies a uniform global curvature. In this case, for $k > 0$, the universe is finite but with no boundary.

For a 3D space CP implies flatness. This is the result of deformation being a local attribute only. In other words, the only "global" curvature possible is zero. In this case, for a finite universe which, of necessity, has a boundary, CP cannot hold true close to the boundary.

The Euclidian nature of our universe is thus not accidental – it is a result of space being 3D and the CP.

Note that expansion of a foamy space is the enlargement of its cells. The number of space cells in the universe is thus considered conserved.

## 5. Notes on Gravitation and Gravitational Field Energy (GFE) Density

### 5.1 Gravitation is the Contraction of Space due to the Presence of a Mass

GR shows that space is curved positively around a mass. This curving is the <u>contraction</u> of space around the mass. Space close to the mass is more contracted than that at a distance, and hence the elastic <u>positive</u> deformation of space. Length, close to a mass, is smaller, and the "running of time" is slower, than at a distance. By positive deformation we mean the 3D analog of positive curving, Section 4.2.

Gravitation is the elastic deformation of space, remove the mass and the deformation is gone.

Space contraction by a mass can be represented by a Gravitational Scale Factor (GSF), as explained below.

Point P, in empty and static space, is a distance r from a point $P_1$ and a distance r + dr from a point $P_2$ on the same line. Introducing a mass M at point P contracts space around it (3D elastic space). The distances from P are now $r' < r$ to $P_1$ and $(r' + dr') < (r + dr)$ to $P_2$ where $dr' < dr$. From the Schwarzschild metric (W. Rindler, 2001 p. 232) (J. Foster and T. A. Nightingale, 2001):

$$dr' = \left(1 - \frac{2GM}{rc^2}\right)^{\frac{1}{2}} dr .$$

For $\frac{2GM}{rc^2} \ll 1$, as for the surface of the sun and the edge of our galaxy, where $\frac{2GM}{rc^2} \sim 10^{-6}$ we get $dr' = \left(1 - \frac{GM}{rc^2}\right) dr$. We define GSF as:

$$(22) \quad a_g(r) \stackrel{\text{def}}{=} \frac{dr'}{dr} = \left(1 - \frac{GM}{rc^2}\right) = \left(1 + \frac{\varphi}{c^2}\right)$$

where φ is the gravitational potential. Thus $a_g$ at the surface of the sun, or at the edge of our galaxy, is approximately $1 - 10^{-6}$ whereas in the last 11 BY the CSF used in cosmology changed, due to expansion, from 0.25 to 1.

From (22) we get the relation of φ to space contraction: $\varphi = c^2 (a_g(r) - 1)$

The gravitational field strength expresses a gradient in space contraction:

$$E_g = \frac{d\varphi}{dr} = c^2 \frac{d}{dr} a_g(r)$$

In the expressions for φ and $E_g$, we ignore the dependence of c on space contraction.

Note that G, the universal gravitational constant tell us by how much a mass M contracts space at a distance r from it.

The CSF, $a_g(r)$, gives the connection, via the metric, between the curvature in the equations and the elastic deformation in reality.

**5.2    Gravitational Field Energy (GFE) Density**

In GR there is no well-defined gravitational field energy (L. D. Landau and E. M. Lifshitz 1962). Standard methods to obtain the energy-momentum tensor yield a non-unique pseudo-tensor. Adding a pseudo-tensor destroys covariance and conservation of energy. It is also asserted, based on the equivalence principle that the gravitational energy cannot be localized (C. W. Misner et al, 1970). Note that in a free-falling frame of reference the gravitational field is nullified (approximately, since a homogeneous gravitational field does not exist) and hence $\in_g$, the GFE density is zero. However, Einstein (1987–2005) himself granted the principle of General Covariance (J. D. Norton, 1993) no more physical meaning than that of a formal heuristic concept.

In the following discussion, $\epsilon_g$ relates to the cosmological frame of reference in which the CMB is isotropic.

In the same way as in EM, the self-energy of a continuous mass distribution is:

$$(23) \quad U(\mathbf{r}) = -\frac{1}{8\pi G}\int_\tau (\nabla\varphi)^2 d\tau + \int_\tau m\varphi d\tau$$

The first term on the right in (23) expresses the GFE, where we interpret the integrand:

$$(24) \quad \epsilon_g = -\frac{1}{8\pi G}(\nabla\varphi)^2 \qquad (\nabla\varphi = \mathbf{E}_g)$$

as the GFE density, and $m\varphi$ as the energy of interaction, or the self-energy of the mass density in the gravitational field. Equation (24) is also the result of GR, for a static weak field (J. Katz, 2008).

GFE has a negative sign since it expresses the work done by the field bringing together material particles.

We consider $\epsilon_g$ to be the inward contractual pressure of space, hence positive, whereas $\epsilon_{CMB}$, as we show, is the outward dilational pressure of space, hence negative. Thus, <u>by their nature</u>, $\epsilon_g$ and $\epsilon_{CMB}$ can be compared.

## 6. Note on CMB Energy Density in an Expanding Universe

Friedmann equations (Rindler 2004 p. 393 equation 18.11) give:

$$(25) \quad \frac{d}{dt}(\rho c^2 a^3) + p\frac{d}{dt}(a^3) = 0$$

A small expanding volume V of space, a ball for example, is proportional to $a^3$. Hence we can replace $a^3$ by V.

Since the energy is $\rho c^2 V = \epsilon V = U$ we get:

$(26) \quad dU + pdV = 0$ which is the equation of continuity (energy balance) in the absence of thermal flow. Isotropy implies no thermal flow and hence the expansion is adiabatic. For space obeying a simple equation of state $p = \omega c^2 \rho$, (26) becomes:

$$(27) \quad \frac{\dot\rho}{\rho} = -(1+\omega)\frac{(a^3)^\cdot}{a^3} \quad \text{Which integrates to:}$$

$(28) \quad \rho \propto a^{-3(1+\omega)}$ For pure radiation $\omega = 1/3$ so that:

$(29) \quad \rho \propto a^{-4}$

This $a^{-4}$ dependence is also considered as the result of photons being red-shifted by space expansion. However, the red shifting does not imply the non-conservation of energy if the pressure of radiation does the work of expansion. Such an interpretation is possible if photons are considered merely as wavepackets of the EM field and not as individual independent particles (R. Loudon, 2000). We conclude that the work done by $\in_R$ to dilate space, by expanding it, is the positive work done to oppose the contraction of space by matter, doing negative work.

Radiation density, like the $\in_{CMB}$, which is homogeneous throughout space, including the interiors of "DM halos" (Granitt et al, 2008) is reduced with expansion. For all $t_2 > t_1$, where $a(t_2) > a(t_1)$, $\in_{CMB}(t_2)$ is related to $\in_{CMB}(t_1)$, as follows:

$$(30) \quad \in_{CMB}(t_2) = \in_{CMB}(t_1)\left(\frac{a(t_1)}{a(t_2)}\right)^4$$

## 7. Notes on Gauss Theorem and the Newtonian Gravitational Field Equation for a Deformed Space

Newtonian gravitation is an approximation since, unlike GR, it does not take into account, in the calculation of the gravitational flux density, the deformation (curving) of space around a mass. However, neither theory considers the inhomogeneous strong deformation of space around a mass due to space expansion. Here we derive the flux density, which is the gravitational field, for this case.

### 7.1 The Extended Gauss Theorem for a Deformed Space

The Gauss Theorem is:

$$(31) \quad \Phi = \int_\sigma \mathbf{E} \cdot \mathbf{d\sigma} = \int_\tau \nabla \cdot \mathbf{E} d\tau$$

where $\mathbf{E}$ is a vector field created by a source – mass, in our case. $\Phi$ is the flux of $\mathbf{E}$ (flux density) through the closed surface $\sigma$ and $\nabla \cdot \mathbf{E} = 4\pi G m$. Consider an expanding space in which $\sigma$ becomes a larger surface, $\sigma'$, and $\tau$ becomes a larger volume $\tau'$, as measured with a rigid yardstick.

$\Phi$ remains the same, hence: $\Phi = \int_\sigma \mathbf{E}' \cdot \mathbf{d\sigma}' = \int_{\tau'} \nabla \cdot \mathbf{E}' d\tau'$

However, formally, $d\sigma \to d\sigma'$ and $d\tau \to d\tau'$ exactly like $\sigma \to \sigma'$ and $\tau \to \tau'$. The conservation of $\Phi$ thus implies that $\mathbf{E}' = \mathbf{E}$ and $\nabla \cdot \mathbf{E}' = \nabla \cdot \mathbf{E}$.

We also get m' = m, since $\int_\tau \nabla \cdot \mathbf{E} d\tau = 4\pi G \int_\tau m d\tau$. The above results are understood by modeling space as being foamy and considering tension, related to permittivity and permeability, in a space cell to be proportional to its linear dimension. In this model, expansion is the enlargement of space cells. Thus $\sigma$ contains the same number of space cells as $\sigma'$ and $\tau$ as $\tau'$. Instead of considering space cells, we can simply refer to a <u>flexible</u> yardstick that is deformed like the space in its location. The result that $\mathbf{E}' = \mathbf{E}$ is surprising. This means that not only the flux $\Phi$ is conserved but also its density, $\mathbf{E}$. This, and the constancy of the velocity of light in space whether expanding or not, (excluding the case of contraction by masses) have the same explanation in the foamy model. After all, light is an EM transverse propagating wave in which the electrical vector field $\mathbf{E}$ is oscillating. The number of cells in an extended space, along a rigid yardstick, is smaller than the number in an un-extended space. However, the overall tension of these cells (like springs in parallel) is the same.

To summarize:

(32) $\quad \int_{\sigma'} \mathbf{E} \cdot \mathbf{d\sigma'} = \int_{\sigma} \mathbf{E} d\sigma \quad$ and $\quad \int_{\tau'} \nabla \cdot \mathbf{E} d\tau' = \int_{\tau} \nabla \cdot \mathbf{E} d\tau$

This, together with (31) is our **Extended Gauss Theorem**.

Note that for the case in which the GSF, $a_g$, has the same order of magnitude as the CSF, a, (32) is only an approximation. A typical $a_g$ is $1 - 10^{-6}$, see Section 5.1, whereas the range of the CSF a that is relevant to our discussion is 0.1 to 1.

## 7.2 The Extended Newtonian Gravitation Field Equation for a Deformed Space

The field strength **E_g**, at a point p, a distance r, from a mass M is $E_g = GM/r^2$. Now consider the expansion of space, whether homogeneously or inhomogeneously, that moves the point p away from M to a distance $r'$, where $r' > r$, as measured with a rigid yardstick. From the above discussion we conclude that:

$$E_g = \frac{\Phi}{A} = E'_g = \frac{\Phi}{A'}$$

where A and A′ are measured by a flexible yardstick, or equivalently, by the number of space cells. Note that A = A′ since space expansion is the enlargement of its cells.

For simplicity, without loss of generality, consider a spherical shell around M with a radius r and scale factor a(r,t) measured by a rigid yardstick at time t. Space expands and the spherical shell, now at $t'$, has a radius $r'$ and a scale factor $a'(r',t')$. In this situation:

$$(\text{number of space cells in the spherical shell}) \propto \frac{4\pi(r')^2}{(a')^2} = \frac{4\pi r^2}{a^2}$$

and hence: $r^2 = \left(\frac{a}{a'}\right)^2 (r')^2$ which gives:

(33) $\quad E_g = E'_g = \frac{GM}{r^2} = \frac{GM}{(r')^2}\left(\frac{a'}{a}\right)^2$

This is our **Extended Newtonian Gravitational Field Equation**.

Note that $E_g$ in (33) appears to be created by a virtual mass:

(34) $\quad M' = M\left[\frac{a'}{a}\right]^2 \quad$ as if $\quad M' = M + M_{\text{Dark Matter}}$.

## 8. Dark Matter Halos are Merely Non-expanding Zones of Space - MZs

In MZs space is strongly deformed and expansion is inhibited (Section 2.2). We show that these MZs mimic the DM Halos.

### 8.1 In and Around Galaxies, Space is Strongly Deformed

To derive the dynamic and kinematic relations that govern the motions of celestial bodies in a galaxy we consider a very simplified model. In this model a galaxy is a "point" mass whose formation time (the mass accretion phase) is much shorter than its present age. In other words, we assume that the galaxy was formed "instantly" at time $t_0$, when the scale factor was $a_0$, possessing its final mass value. Note that in this section $a_0$ is the scale factor value at the time of the galaxy formation and not the present value. The redshifted galactic light recorded now left the galaxy at cosmic time $t_z$, when the scale factor was $a_z$. We divide the space around a galaxy into three regions according to the relative values of $|\epsilon_g|$ and $\epsilon_{CMB}$:

a. **From the center of a galaxy to $R_0$, where $|\epsilon_g| \geq \epsilon_{CMB}$**

   $R_0$ is the distance at which, initially, at the time of formation $|\epsilon_g| = \epsilon_{CMB}$. In this region, the local contraction of space by the mass of the galaxy is stronger than the opposing dilation caused by the CMB. Space expansion is inhibited in this region, as explained in Section 2.2, and hence Newtonian gravitation is applicable.

b. **From $R_0$ to R**
   R is the distance for which $|\epsilon_g|$ was equal to $\epsilon_{CMB}$ at the time of emission of a photon that reaches us now.
   In this region, equilibrium was first attained at a distance $R_0$, at the time, $t_0$, of formation of the galaxy. The expansion of the surrounding space beyond $R_0$, due to the expansion of the universe, lowered the $\epsilon_{CMB}$, and hence equilibrium was reached for $t > t_0$, at a greater distance $r(t) > R_0$. This is an ongoing process in which the region surrounding $R_0$ grows with time, with an ever-increasing value of the scale factor.
   Light that reaches us now, left the galaxy at time $t_z$. Equilibrium, $|\epsilon_g| = \epsilon_{CMB}$, at this time, occurred at a distance R from the center of the galaxy. Space density in the region between $R_0$ and R is "frozen", since $|\epsilon_g| > \epsilon_{CMB}$. Space density at $R_0$ is larger than at R. In this region, RCs are flat, as our Extended Newtonian Gravitational Law predicts, see Section 8.2. This region is the "DM Halo".

c. **From R onwards**
   In this region, where $|\epsilon_g| < \epsilon_{CMB}$, space expands freely.

To express this space inhomogeneity, we consider the CSF in the region between $R_0$ and R to be dependent not only on time, t, but also on the distance, r, from the center of a galaxy, thus the scale factor a is a(r,t).

## 8.2 The Gravitational Central Acceleration, in and around Galaxies for the Region between $R_0$ and $R$

Consider a point in the second region, $R_0 < r < R$. We derive $\epsilon_g(r,t)$ in this region by using the extended gravitational field equation (33) with the following notational changes:

$R_0$ instead of r

r instead of r'

$a = a(R_0,t_0)$ but since a, at $R_0$, does not change with time for all $t > t_0$, $a = a(R_0,t)$ and $a' = a(r,t)$. With these changes (33) becomes:

$$E_g(r,t) = \frac{GM}{r^2} \left[ \frac{a(r,t)}{a(R_0,t)} \right]^2$$

Substituting $E_g(r,t)$ in equation (24) for $|\epsilon_g|$ gives:

$$(35) \quad |\epsilon_g(r,t)| = \frac{1}{8\pi G} E_g^2(r,t) = \frac{1}{8\pi G} \left[ \frac{GM}{r^2} \cdot \left[ \frac{a(r,t)}{a(R_0,t_0)} \right]^2 \right]^2$$

Equation (30), with the above notational changes, is:

$$(36) \quad \epsilon_{CMB}(t) = \epsilon_{CMB}(t_0) \cdot \left[ \frac{a(r,t)}{a(R_0,t_0)} \right]^{-4}$$

In the region $R_0 < r < R$, $|\epsilon_g| = \epsilon_{CMB}$, hence, equating (36) to (35) gives:

$$(37) \quad \frac{a(r,t)}{a(R_0,t)} = \left[ \frac{8\pi G \, \epsilon_{CMB}(t_0)}{G^2 M^2} \right]^{\frac{1}{8}} \cdot r^{\frac{1}{2}}$$

We designate $E_g$ by g and the numerator in (37) by:

$$(38) \quad g_0^2 = 8\pi G \, \epsilon_{CMB}(t_0)$$

This designation is explained at the end of this section and in the following section.

We rewrite (37) as:

$$(39) \quad \frac{a(r,t)}{a(R_0,t_0)} = \left[ \frac{g_0^2}{G^2 M^2} \right]^{\frac{1}{8}} \cdot r^{\frac{1}{2}}$$

Substituting (39) into (33) gives:

$$(40) \quad g = \frac{GM}{r^2} \cdot \left[ \frac{g_0^2}{G^2 M^2} \right]^{\frac{1}{4}} \cdot r = \left[ (GM)^2 \cdot \frac{g_0}{GM} \right]^{\frac{1}{2}} \cdot r^{-1} = [(GM) \cdot g_0]^{\frac{1}{2}} \cdot r^{-1}$$

Thus the gravitational central acceleration in the region R to $R_0$ is:

(41) $\quad g = \dfrac{\sqrt{g_0 GM}}{r}$ $\quad\quad$ which resembles the **Milgrom** (1983) relation, but is in no way related to the MOND paradigm.

Squaring equation (41), gives $g^2/g_0 = GM/r^2$

Since $g = \dfrac{v^2}{r}$ we get: $\dfrac{v^4}{r^2} = g_0 \dfrac{GM}{r^2}$ $\quad$ or:

(42) $\quad v^4 = (g_0 G)M = B \cdot M$ $\quad\quad$ which is the **Tully-Fisher** relation.

$\quad\quad\quad B = (g_0 G)$

The circular rotation velocity in this region is:

(43) $\quad v = (g_0 GM)^{\frac{1}{4}}$

and thus <u>RCs in this region are flat</u>. Section 9 shows that a more realistic model that takes into account the evolution of galaxies yields RCs that fit observed RCs.

From equation (38) that defines $g_0$:

(44) $\quad \dfrac{1}{8\pi G} g_0^2 = \epsilon_{CMB}(t_0)$

Thus, $g_0$ is the field strength (central acceleration) at $R_0$, at the time, $t_0$, of formation.

Note that the region, $R_0$ to R, in which space density is frozen, grows with time. At $R_0$ space density is high – small $a(r,t)$ – and is reduced towards R – higher $a(r,t)$. At distances $r > R$, where $\epsilon_{CMB} > \epsilon_g$, space expands.

### 8.3 Some Numerical Results for $g_0$, B, $R_0$ and R

- **For $g_0$**
  For galaxies formed at $z \sim 3$ the corresponding scale factor, a, is 0.25, (Baugh et al, 1998) and hence the time of formation, $t_0$, is ~11 BY. To obtain the value for $\epsilon_{CMB}(t_0)$ we use equation (30).
  The present value, $\epsilon_{CMB}(Now) = 4.17 \cdot 10^{-13}$ erg cm$^{-3}$, gives for $t_0$:
  $\epsilon_{CMB}(t_0) \sim 1.0 \cdot 10^{-10}$ erg cm$^{-3}$. Hence, from (38) we get:
  (45) $\quad g_0 = \sqrt{8\pi G \epsilon_{CMB}(t_0)} \sim 1.3 \cdot 10^{-8}$ cm s$^{-2}$.
  This result is close to the Milgrom (1983) "universal constant" but is not a constant at all (B. Famaey et al, 2007). Observations show that the central acceleration $g_0$ takes a wide range of values, as our expression for this parameter predicts, see Begeman et al (1991) and Scott et al (2001).

- **For B, the Tully-Fisher Parameter**
  In a large sample of galaxies, covering a wide dynamic range, McGaugh et al (2000) found the Tully-Fisher relation $M = AV^4$ between the baryonic mass, M, of galaxies and the rotation velocities at the flat edge of their RCs. The coefficient A in this relation is found empirically to be $A = 35 h_{75}^{-1} M(Sun) s^{-4} km^{-4}$. Here $h_{75}$ is the Hubble constant in units of 75 (km/s$^{-1}$)/Mpc. With h = 1, A = 7 x 10$^{14}$ gr s$^{4}$ cm$^{-4}$. We write the Tully-Fisher relation, (72), as $V^4 = BM$, where according to (68) $B = G\sqrt{8\pi G \epsilon_{CMB}(t_0)}$, for $t_0$, the time of the galaxy formation. Clearly B = 1/A.

Using (30) $\epsilon_{CMB}(t_0) = a^{-4}(t_0) \epsilon_{CMB}(Now) = (1+z_0)^4 \epsilon_{CMB}(Now)$ we obtain a fit of our B to the value 1/A for $z_0 = 4$. Baugh et al (1998) derived the value z = 3 to 3.5, for the redshift at the time of galaxy formation, from an entirely independent set of observations. Our formation time differs from that of Baugh, which is the epoch at which a galaxy first becomes detectable in optical and IR light. It is likely that this time precedes the birth of light emitting objects in a proto-galaxy. This may explain the difference between the two results.

- **For $R_0$**

  For a given M the distance $R_0$ is:

  (46) $\quad R_0 = \sqrt{\dfrac{GM}{g_0}}$

  In our simplified model, this is the distance from the center of a galaxy at which the "DM halo" starts. As an example, for a galaxy formed ~11 BY ago with a bulge mass $M \sim 1.3 \cdot 10^{10} M_\odot$, our calculation gives $R_0 \sim 3$ KPC. Assuming similar initial conditions for the Milky Way galaxy, the above calculation of $R_0$ is supported by observations, (Gerhard, 2002). From this distance onwards, the rotational velocity increases, reaches a "plateau" and then decreases, as is indicated by observations of dispersion velocities (Battagalia, 2005).

- **For R**

  From equation (41) $R = \dfrac{\sqrt{GM}}{g}$.

  $g(today) = \sqrt{8\pi G \epsilon_{CMB}(today)} \sim 8 \cdot 10^{-10}$ cm s$^{-2}$

  For the Milky Way galaxy, $R \sim 10^{23}$ cm ~ 100 KPC.

  Since $\epsilon_{CMB} = \epsilon_{CMB}(today) \cdot a^{-4}$ we get $R \propto a$.

To summarize, in our simplified model of a galaxy with mass M, the gravitational field, $E_g = g$, around a galaxy, for the three regions of an RC, is:

(47) $\quad r \leq R_0 \qquad\qquad g = \dfrac{GM}{r^2}$

(48) $\quad R_0 < r \leq R \qquad g = \dfrac{\sqrt{g_0 GM}}{r}$

In reality, g in the first and second regions depends on the mass distribution and the history of the galaxy formation.

(49) $\quad r > R \quad\quad\quad g = \dfrac{GM}{r^2}\left[\dfrac{a_z}{a_0}\right]^2$

$a_0$ the scale factor at the time of formation of the galaxy.
$a_z$ the scale factor at the time of emission of light towards us that reaches us now.

Equation (49) gives $g_{at\,R} = \dfrac{GM}{R^2}\left[\dfrac{a_z}{a_0}\right]^2$. In the adjacent region (48) gives $g_{at\,R} = \dfrac{\sqrt{g_o GM}}{R}$.

Equating these expressions gives $R = \sqrt{\dfrac{GM}{g_0}}\cdot\left[\dfrac{a_z}{a_0}\right]^2$. Since (46) is $\sqrt{\dfrac{GM}{g_0}} = R_0$, we get:

(50) $\quad R = R_0\left[\dfrac{a_z}{a_0}\right]^2$

With time, the zone of flat RCs grows. This means that with time "DM halos" should grow. Observations confirm this result (Massey et al, 2007).

## 9 Galactic Evolution is Taken into Account in a Model that Yields Realistic RCs

### 9.1 The Model

Galaxies attain their observed baryonic masses and mass distribution during a time span that is a fraction of the age of the universe (Searle and Zinn 1978, Martinez-Delgao et al 2008). The evolution of a deformed space halo around a galaxy is determined by both the decline of $\in_{CMB}$ with space expansion, and the history of the accumulation or loss of mass by the galaxy.

We build a model (Appendix B) that takes this history into account.

### 9.2 Our Theoretical RCs Compared with Observed RCs

We show three theoretical RCs, obtained from the model developed in Appendix B, with no need for DM and compare them qualitatively with observed RCs of three galaxies.

In our simple model, an initial accretion phase is followed by one or two phases of mass accretion or loss (due, for example, to SN explosions or stellar winds). In this model, mass accretion or loss occurs at a constant rate. The duration of each phase, and the accretion or loss rates, are free parameters of the model.

The large frames on the left-hand side of Figure 2 present the observed RCs of the galaxies NGC 2903, NGC 3657 and UGC 4458. (de Blok et al 2008, Milgrom 2008, and Sanders and Noordermeer 2007, respectively).

The thick lines in the curves on the right-hand side are the theoretical model curves. The thin lines are the corresponding Newtonian curves. The x-axis is the normalized radial distance ρ, as defined above. The numbers on the y-axis are dimensionless, expressing rotation velocities in units of $\sqrt{GM_0/R_0}$.

The curves below each RC plot show the assumed evolution of the models used to generate the theoretical curves. They show for each model the (normalized) mass of the galaxy as a function of cosmic time t (left curve), taking as unity the present age of the universe, T = 1, and redshift z (right curve).

The theoretical RCs fit the observed RCs.

The similarity in the profile of the theoretical plots to that of the observed RCs is evident. However, we do not claim that the parameters of our model necessarily characterize the true histories of the three galaxies. They are not even determined uniquely by the profiles of the curves alone. As discussed above, our model is over simplistic and does not take into account observed data of the real galaxies such as their surface brightness. The purpose of this exposition is merely to demonstrate qualitatively that our theory is capable of explaining observed RCs of galaxies, even at very large distances from their centers, with no need of any mass in addition to that of the observed luminous baryonic matter. To establish the fit on a more quantitative footing, much more work is required to develop equations of non-spherical mass distributions, which should also incorporate data of measurements in real galaxies.

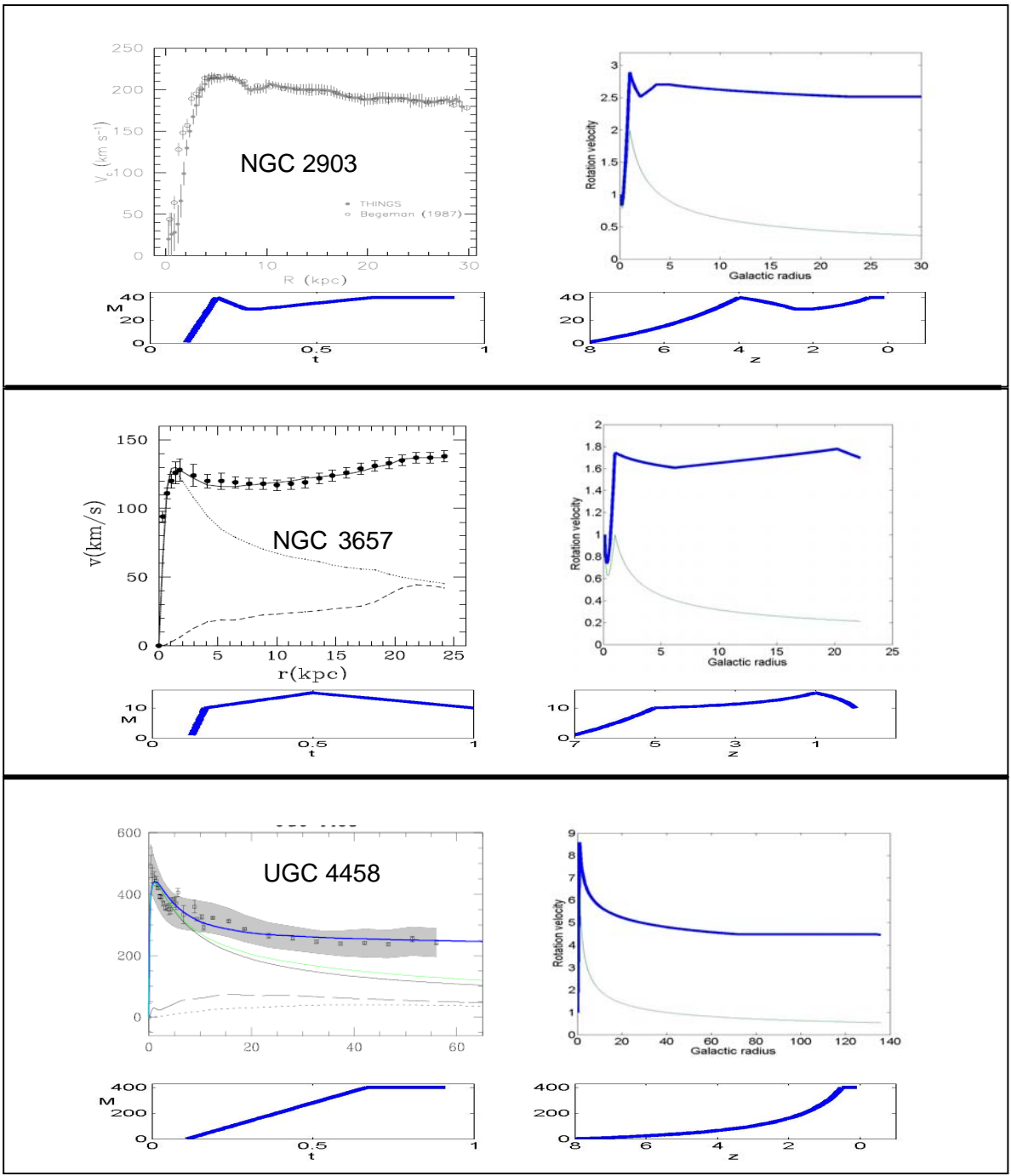

**Figure 2. Comparison of our Theoretical RC with Observed RCs**

## 10. Notes on Phenomena in MZs

### 10.1 The Gravitational Potential is Modified by Space Expansion

By integrating equation (41) for g, we get the potential difference.

$$\varphi(r) - \varphi(R_0) = \int_{R_0}^{r} g\, dr = \sqrt{g_0(GM)} \cdot \int_{R_0}^{r} r^{-1} dr = \sqrt{g_0(GM)} \cdot \ln\frac{r}{R_0}$$

This potential difference is only valid in the region between $R_0$ and R. Since:
$$\varphi(R_0) = -GM/R_0 \quad \text{(in reality, } \varphi(R_0) \text{ depends on the mass distribution) we get:}$$

(56) $\quad \varphi(r) = \sqrt{g_0(GM)} \cdot \ln\frac{r}{R_0} - \frac{GM}{R_0}$

### 10.2 The Gravitational Potential in an Expanding Universe Explains the Enhanced Gravitation Lensing

A point mass, M, which serves as a lens, deflects a light beam with an impact parameter, b, at the following deflection angle:

(57) $\quad \alpha = \frac{4GM}{c^2 b} = \frac{4}{c^2} \varphi$

where $\varphi$ is the gravitational potential at a distance b from M (Carroll, 2004, Sec. 7.3 and Sec. 8.6).

However, the potential in the zone of flat RCs around M, expressed by (51), yields, for large impact parameters, a much larger deflection of light beams.

### 10.3 "DM Halos" can be Detached From Matter

"DM halos" can be detached from fast moving matter, as evidenced by the "bullet cluster" 1E0657-56, (Clow et al, 2004). We thus conclude that mass deforms space in two ways:

- **Elastic deformation by the presence of mass alone**
  GR states that space deformation is gravity, i.e., in the vicinity of masses, space is contracted. This contraction is elastic - remove the mass and space resumes its original geometry.

- **Non-elastic deformation due to space expansion around a mass**
  In addition to the above elastic deformation, space is also deformed by the inhomogeneous space expansion around the mass, caused by the mass. Such deformation is observed as a "DM halo", as shown above. However, in contrast to elastic deformation, <u>the halo does not follow a moving mass and retains its geometry</u>, as if the mass is still in its original location. Note that elastic deformation is orders of magnitude smaller than the non-elastic deformation (Section 5.1) accumulated over cosmological time.

## 11.  Summary

We have dispelled the need for Dark Energy and Dark Matter.

To dispel the need for Dark Energy we have applied General Relativity to the empty inter-galactic zones of space in our 3D universe, rather than to its entirety.

To dispel the need for Dark Matter, we have related to the strong space deformation, in and around galaxies, caused by the interplay between the gravitational field energy density (positive pressure) and the Cosmological Microwave Background energy density, (which, unconventionally, we consider a negative pressure). By developing an extended Gauss theorem for deformed space, we have obtained the dynamic and kinematic laws of motion for celestial bodies moving in and around the centers of galaxies.

Our theoretical work has resulted in a fit to observations. For Dark Energy, we have obtained a fit to data from Supernovae, and for Dark Matter a fit to observed Rotation Curves.

## Acknowledgments

We would like to thank Roger M. Kaye for his linguistic contribution and technical assistance.

## Appendix A (The Corrected LD – Section 3.2)

A slice of space between z and z + dz corresponds to time duration between t and dt. The frozen halo grows by dr, where $dr = \frac{dr}{dt} dt$.

Equation (5), $\dot{a} = const$, gives $a(t) = a(T) + (t-T) \cdot \dot{a}(t)$.

Since $\frac{\dot{a}(T)}{a(T)} = H_0$, $a(z) = \frac{1}{1+z}$ and $T - t = \frac{1}{H_0} \cdot \frac{z}{1+z}$ we get $dt = -\frac{1}{H_0} \cdot \frac{dz}{(1+z)^2}$.

From (15) we get $\frac{dr}{dt} = \frac{r_0}{a_0^2} 2 a(t) \cdot \dot{a}(t)$ and since $a(t) = a(z) = \frac{1}{1+z}$ we get:

$$dr = -2 \frac{r_0 \dot{a}}{a_0^2} \cdot \frac{1}{(1+z)} \cdot \frac{1}{H_0} \cdot \frac{dz}{(1+z)^2} = -2 \frac{r_0 \dot{a}}{a_0^2} \cdot \frac{1}{H_0} \cdot \frac{dz}{(1+z)^3} \quad \text{but} \quad \dot{a}(T) = H_0, \text{ hence:}$$

(17) $\quad D_M(z) = -2 \frac{r_0}{a_0^2} \int_z^0 \frac{dz}{(1+z)^3} = \frac{r_0}{a_0^2} \left[ 1 - \frac{1}{(1+z)^2} \right]$

Extracting $r_0$ as a function of β from (16), and substituting it in (17) gives:

(18) $\quad D_M(z) = \beta R(T) \left[ 1 - \left(\frac{1}{1+z}\right)^2 \right] = \beta \left[ 1 - \left(\frac{1}{1+z}\right)^2 \right] R_I(T) \cdot \left( 1 + \frac{R_M(T)}{R_I(T)} \right)$

However, $\left( \frac{R_M(T)}{R_I(T)} \right) \ll 1$, therefore:

(19) $\quad D_M(z) \cong \beta \left[ 1 - \frac{1}{(1+z)^2} \right] \frac{c}{H_0} \cdot \ln(1+z)$

The expectation value of the distance from a center of a disk of radius R of a random point on the disk surface is:

$$\bar{r} = \int_0^R r p(r) dr = \int_0^R r \frac{2\pi r dr}{\pi R^2} = \frac{2}{R^2} \int_0^R r^2 dr = \frac{2}{3} R$$

Half the length of a cord at a distance 2/3R from the center of a sphere of radius R is $\sqrt{R^2 - 4/9 R^2} = \sqrt{5/9} R \cong 0.75 R$. Therefore, the mean distance (the cord) within the frozen sphere of radius r(t) is 1.5 r(t).

As a result of the above calculations, we get for the corrected geometric LD:

(20) $\quad D_L = \frac{c}{H_0} \left\{ 1 - 1.5 \beta \left[ 1 - \frac{1}{(1+z)^2} \right] \ln(1+z) \right\}$

The corrected LD for the web-like universe is: $d_L = (1+z) \cdot D_L$, which gives the corrected distance modulus equation (21).

## Appendix B (The Model – Section 9.1)

To account for this history we introduce the following functions and parameters:

- $\mu(t) = M(t)/M_0$ describes the mass evolution of the galaxy, normalized to the galactic mass at formation.

- $\chi(r) = M(T,r)/M_0$ represents the mass distribution in a finalized galaxy, as observed. Specifically, $\chi(r)$ is the (normalized) mass of a sphere of radius r in the observed galaxy. For $r > R_T$, where $R_T$ is the radius of the spherical distribution of the baryonic matter of the mature galaxy, $\chi(r) = \mu_T$. Here $\mu_T = M_T/M_0$ is the observed (normalized) mass of the galaxy.

- $\xi = R_T/R_0$ is the ratio of $R_T$, the radius of the mature galaxy, to $R_0$, the radius of the infant galaxy.

- $\rho = r/R_T$ expresses distances from the center of a galaxy with a dimensionless normalized radial coordinate

Our simplifying assumptions are:

- After the formation of the galaxy, at time $t_0$, all accreted matter is distributed instantly according to the observed final distribution.
- The radius of the frozen sphere of space is always larger than the instantaneous radius of the galaxy.

This model, using equations (39) and (46), and designating $a(R_0, t_0) = a_0$, gives:

$$(51) \quad \frac{r(t)}{R_0} = \mu^{\frac{1}{2}}(t) \left[\frac{a(t)}{a_0}\right]^2.$$

This can also be written for a piecewise linear $\mu(t)$ as a function of a:

$$(52) \quad \mu^{\frac{1}{2}}(a) \left(\frac{a}{a_0}\right)^2 - \rho = 0$$

We now compute the gravitational field strength at every radius $\rho$, $\rho_0 \leq \rho \leq \rho_z$ as follows:

For each $\rho$ we consider expression (52) as an equation for $a = a(\rho)$, recalling that $a_0 = (1+z_0)^{-1}$. Its solution is the cosmic scale factor, a, at which the radius of the frozen sphere arrives at the distance $\rho$ from the center. From that moment on, space expansion is frozen at this point with this value a as the local scale factor.

Substituting this value, $a(\rho)$, in equation (33), where $E_g = g$, and from the equation for the circular rotation velocity, $v = \sqrt{rg}$, we obtain:

$$(53) \quad v(\rho) = \sqrt{\frac{GM_0}{R_0}} \sqrt{\frac{\chi(\rho)}{\rho}} \left(\frac{a}{a_0}\right).$$

At the outskirts of the galaxy, for $\rho > \rho_z$ where $\rho_z$ is the radius of the frozen sphere at the time the recorded photon left the galaxy, the Extended Newtonian expression gives:

$$(54) \quad v(\rho) = \sqrt{\frac{GM_0}{R_0}} \sqrt{\frac{\chi(\rho)}{\rho}} \left(\frac{a_z}{a_0}\right).$$

Here, as above, $a_z = (1+z)^{-1}$ is the value of the cosmic scale factor for the measured redshift of the galaxy. Rotation velocity, for all $\rho \geq \rho_0$, is:

$$(55) \quad v_N(\rho) = \sqrt{\frac{GM_0}{R_0}} \sqrt{\frac{\chi(\rho)}{\rho}}$$

## List of Acronyms

| | | | |
|---|---|---|---|
| BB | Big Bang | GFE | Gravitation Field Energy |
| BBN | Big Bang Nucleosynthesis | GR | General Relativity |
| BY | Billion Years | GSF | Gravitational Scale Factor |
| CMB | Cosmological Microwave Background | HS | Hubble Sphere |
| CSF | Cosmological Scale Factor | IZ | Inter-galactic Zone |
| DE | Dark Energy | ΛCDM | Λ Cold Dark Matter |
| DM | Dark Matter | KPC | Kiloparsec |
| DM μ | Distance Modulus μ | LD | Luminosity Distance $d_L$ |
| EM | ElectroMagnetic | MZ | Mass Zone |
| EP | Elementary Particle | NL | Non-linear |
| FRW | Friedmann Robertson Walker | QED | Quantum ElectroDynamics |
| GC | General Covariance | RC | Rotation Curve |
| GE | Gravitational Energy | ZPF | Zero Point Fluctuations |